\documentclass[aps,prl,twocolumn,superscriptaddress, amsmath,amssymb]{revtex4-2}

\usepackage{graphicx}% Include figure files
\usepackage[linkcolor=blue]{hyperref}
\usepackage{dcolumn}% Align table columns on decimal point
\usepackage{bm}% bold math
\usepackage{xcolor}
\begin{document}

%\preprint{}

\title{Probing Vibronic Coherence in Charge Migration of Molecules Using Strong Field Sequential Double Ionization}


\author{C. H.  Yuen}
\email[]{iyuen@phys.ksu.edu}
\author{C. D.  Lin}
\email[]{cdlin@phys.ksu.edu}
\affiliation{J. R. Macdonald Laboratory, Department of Physics, Kansas State University, Manhattan, Kansas 66506, USA}
%\thanks{}

%\date{\today}

\begin{abstract}
We propose a novel scheme for probing vibronic coherence in charge migration in molecules utilizing strong field sequential double ionization. 
To demonstrate the feasibility of this approach,  we perform full simulations of a pump-probe scheme employing few-cycle intense infrared pulses for N$_2$ and O$_2$. 
We predict that the vibronic coherence between the pumped states will be directly imprinted in experimental observables such as kinetic energy release spectra and branching ratios of the dissociative dications.
Our simulations are based on the recently developed DM-SDI model,  which is capable of efficiently accounting for molecular orientations and enabling direct comparison with experimental results. 
Our findings strongly encourage the use of this probing scheme in future charge migration experiments.
\end{abstract}

\maketitle

In a pump-probe experiment for studying molecular dynamics,  one strives for a high temporal resolution as well as a large signal-to-noise ratio.
While a high temporal resolution can be achieved using isolated attosecond pulses,  the photon flux from table-top light sources is generally too low. 
Higher count rates and larger signal-to-noise ratios can be reached using intense infrared (IR) pulses. 
If the IR pulse is used as a probe,  then it is desirable to increase its peak intensity for higher count rates.
As the peak intensity increases,  the probe laser may cause sequential double ionization (SDI) of the neutral molecule,  resulting in molecular fragmentation.
Although experimental setups for making intense few-cycle IR laser pulses and coincidence measurement for ion fragments have been widely available~\cite{voss2004, alnaser2004,wu2010b, de2010, de2011, xie2014, cheng2021},
this probing scheme has been unfavorable due to the lack of theoretical support.
Recently,  we developed a density matrix approach for sequential double ionization (DM-SDI) of molecules~\cite{yuen2022,yuen2023},  which was benchmarked with experiments for $\rm N_2$ and $\rm O_2$ at their ground states~\cite{voss2004,  wu2010b}.
One of the goals of this Letter is to demonstrate the feasibility of using SDI to probe molecular dynamics.

The dynamics of interest are the charge migration of molecules.
Charge migration is typically initiated by the removal of an electron,  which left the molecular ion in a superposition of electronic states,  then the electron cloud will migrate along the molecular skeleton~\cite{cederbaum1999,  folorunso2021}. 
Probing this process has become increasingly important due to the advancement of attosecond science~\cite{krausz2009,biegert2021},
with prospects of observing electron motion in a molecule in real-time and ultimately controlling molecular dynamics.
% by timing some actions with respect to the onset of charge migration.
A critical problem in charge migration studies is that the nuclear motion will set in after a few femtoseconds,  and could lead to the decoherence between different electronic states.
Such decoherence would be an obstacle for observing pure electron dynamics,  but it could lead to a permanent charge transfer to a different site in the molecule,  which offers opportunities for controlling its chemical reactivity~\cite{lepine2014}.
While the effects of nuclear motion on charge migration have been studied theoretically~\cite{despre2015, arnold2017,golubev2020,dey2022,scheidegger2022},  it remains challenging to monitor the coherence experimentally.

A necessary condition for observing the coherence is that the superposition of states must reach the same final states after the probing process.
%the observables must have contribution from the interference between formation pathways from the superposition of states. 
However,  it is not a sufficient condition since the interference signals could be too weak to be observed.
This is indeed one of the major challenges in probing the charge migration: It is unclear that,  for what type of molecules and for what type of probing processes,  such interference signals could be detected.
There have been some successful experimental investigations,  for example,  by using attosecond extreme-ultraviolet (XUV) pump and IR probe~\cite{sansone2010,  calegari2014,  lara2018},  high-harmonic spectroscopy~\cite{smirnova2009,kraus2015,He2022},  attosecond transient absorption spectroscopy (ATAS)~\cite{kobayashi2020a,kobayashi2020,matselyukh2022},  attosecond pulse train pump-probe~\cite{okino2015,fukahori2020},  and femtosecond X-ray pump-probe~\cite{barillot2021,schwickert2022}.
But among these experiments,  the types of the target molecules and the probing mechanisms greatly varied,  and it is uncertain whether a particular probing scheme will work on other targets.
A promising scheme in which the observables could be simulated is the ATAS~\cite{santra2011,golubev2021},
but the interference signals may vanish after averaging over molecular orientations due to the anisotropy of the coherence.
% since the coherence between some pumped states is anisotropic.
%but the interference signals may be small compared to the incoherent signals after averaging over molecular orientations.
The search for a general and experimentally accessible probing scheme for charge migration of molecules is therefore of great importance.
%Therefore,  it is of the utmost importance to search for a general and experimentally accessible probing scheme for charge migration of molecules.

%Among the successful experiments,  a particular promising approach to track the decoherence due to nuclear motion is the ATAS.
%The theory behind ATAS has been well understood~\cite{santra2011,golubev2021}.
%The basic mechanism of the probe scheme is to drive the core-valence transition for a pair of pumped states to the same core-hole state using a XUV or a X-ray photon.
%%Since the dipole moment of the molecular ensemble adds up coherently,  interference signal between the pumped states could be observed in the ATAS.
%Consequently,  according to the Wigner-Eckart theorem,  only the coherence between pumped states with similar symmetries could be observed in the ATAS.
%%In the case of homonuclear diatomic molecules,  the theorem implies the pumped states must have the same parity,  such that the states are dipole-forbidden.
%Moreover,  the orientation effects of molecules have not been fully accounted for in previous theoretical treatments,  since the coherence of the pumped states depends on molecular orientation in general.
%Upon orientation averaging,  it is possible that the interference signals to be washed out. 

\begin{figure}
\includegraphics[scale=0.2]{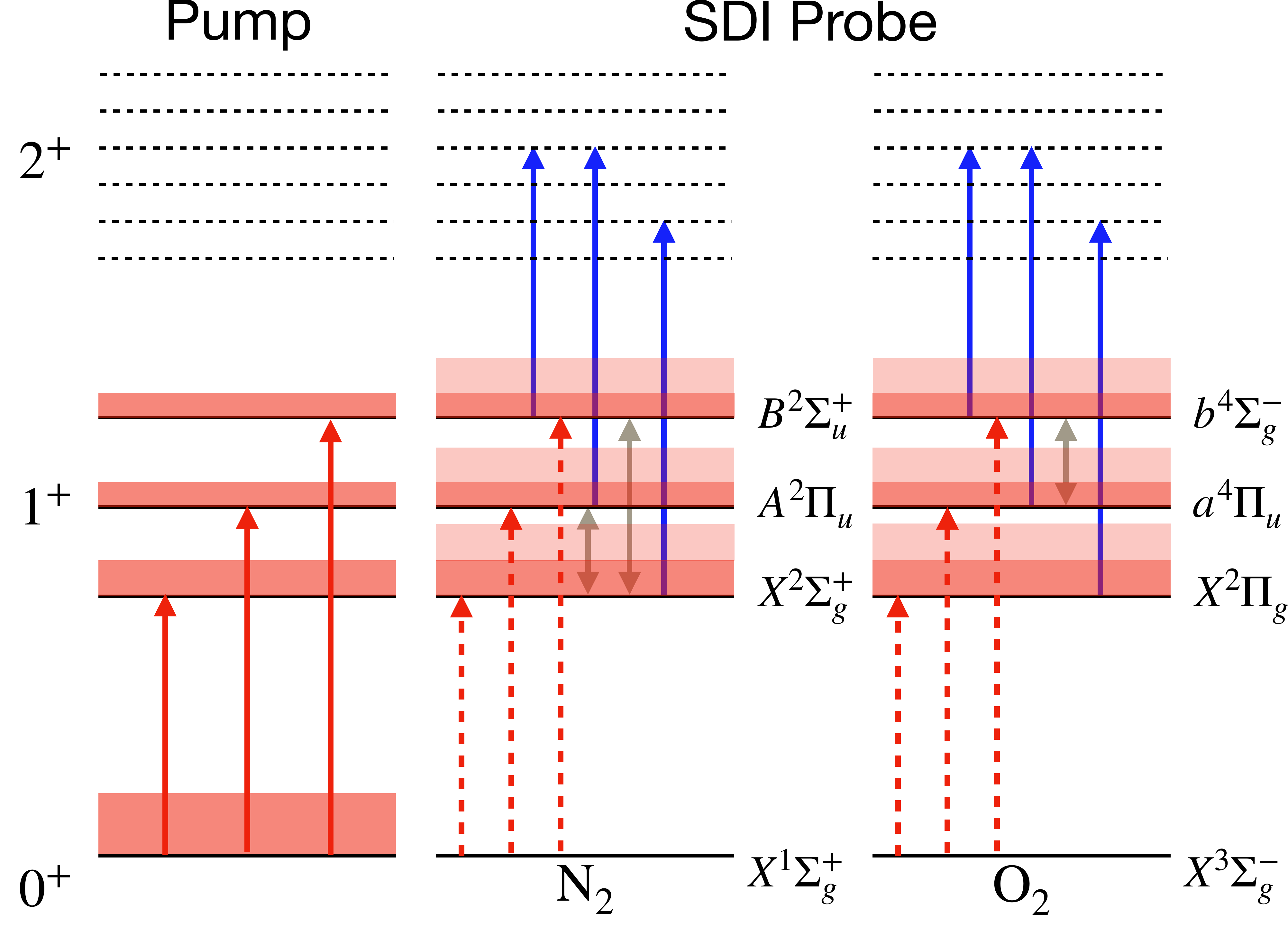}
\caption{Illustration of the charge migration pump-probe scheme for N$_2$ and O$_2$.  The pump laser populates the lowest three ionic states coherently (red solid arrows),  and the populations are represented by the shaded areas.  At a later time,  the intense few-cycle IR probe pulse ionizes the remaining neutral population (red dashed arrows),  couples the ionic states (gray solid arrows),  and tunnel ionizes them to form dications (blue solid arrows).  The coherence between the pumped states influences the dication yield through interference with the nascent ionic states (lightly shaded areas) and the laser couplings.}
\label{figs:illu}
\end{figure}

SDI could be an excellent process for probing coherence in charge migration since it is driven by laser couplings between the ionic states~\cite{yuen2022,yuen2023}.
Suppose a pump laser singly ionizes the neutral molecule and forms a superposition of ionic states.
Then,  the IR probe will create nascent ionic states,  couple the ionic states,  and further ionize them to dications.
The coherence between the pumped states controls the transient ionic populations through interference with the nascent ionic states and the laser couplings.
Since the dication yields depend on the transient ionic populations,  the yields will change with the coherence as well.
As a result,  the coherence could be revealed in experimental observables such as kinetic energy release (KER) spectra or branching ratios of the dications,  which always survive the orientation averaging.
Figure~\ref{figs:illu} illustrates this pump-probe scheme for N$_2$ and O$_2$.
Since the mechanism of SDI should be general for any molecules and the required experimental setups are widely available,  the SDI process can serve as a general probing scheme for charge migration.
%The transient ionic populations depend sensitively on the coherence between the pumped states due to the interference with the nascent ionic states and the laser couplings.
%As the dications yields at different orientations add up incoherently,  the signals always survive the orientation averaging.
%couple the pumped states via their transition dipole moments and further ionize them to dications,  as well as doubly ionize the remaining neutral population.
%The nascent ionic states created from the probe laser will interfere with the pumped ionic states,  and the coherence between the ionic states will strongly affect their populations through the laser couplings.
We note that the SDI probe could be similar to the IR probe used in Refs.~\cite{calegari2014,  lara2018},  where some dication yields were measured.

In this Letter,  we provide compelling evidence on the viability of using the SDI process to probe charge migration in molecules.
This is supported by complete simulations on the pump-probe scheme in Fig.~\ref{figs:illu} for N$_2$ and O$_2$ using the DM-SDI model~\cite{yuen2022,  yuen2023},  which 
takes molecular orientations into full account such that the results can be compared with future experiments directly.
We discover that the vibronic coherence is remarkably imprinted in the KER spectra and branching ratios for $\rm N^+ + N^+$ and $\rm O^+ + O^+$ at different pump-probe delays.
The findings of this Letter strongly encourage future experiments to use the SDI probe for charge migration in N$_2$ and O$_2$ as well as other molecules.

\begin{figure*}[ht]
\centering
\begin{tabular}{cc}
\includegraphics[scale=0.435]{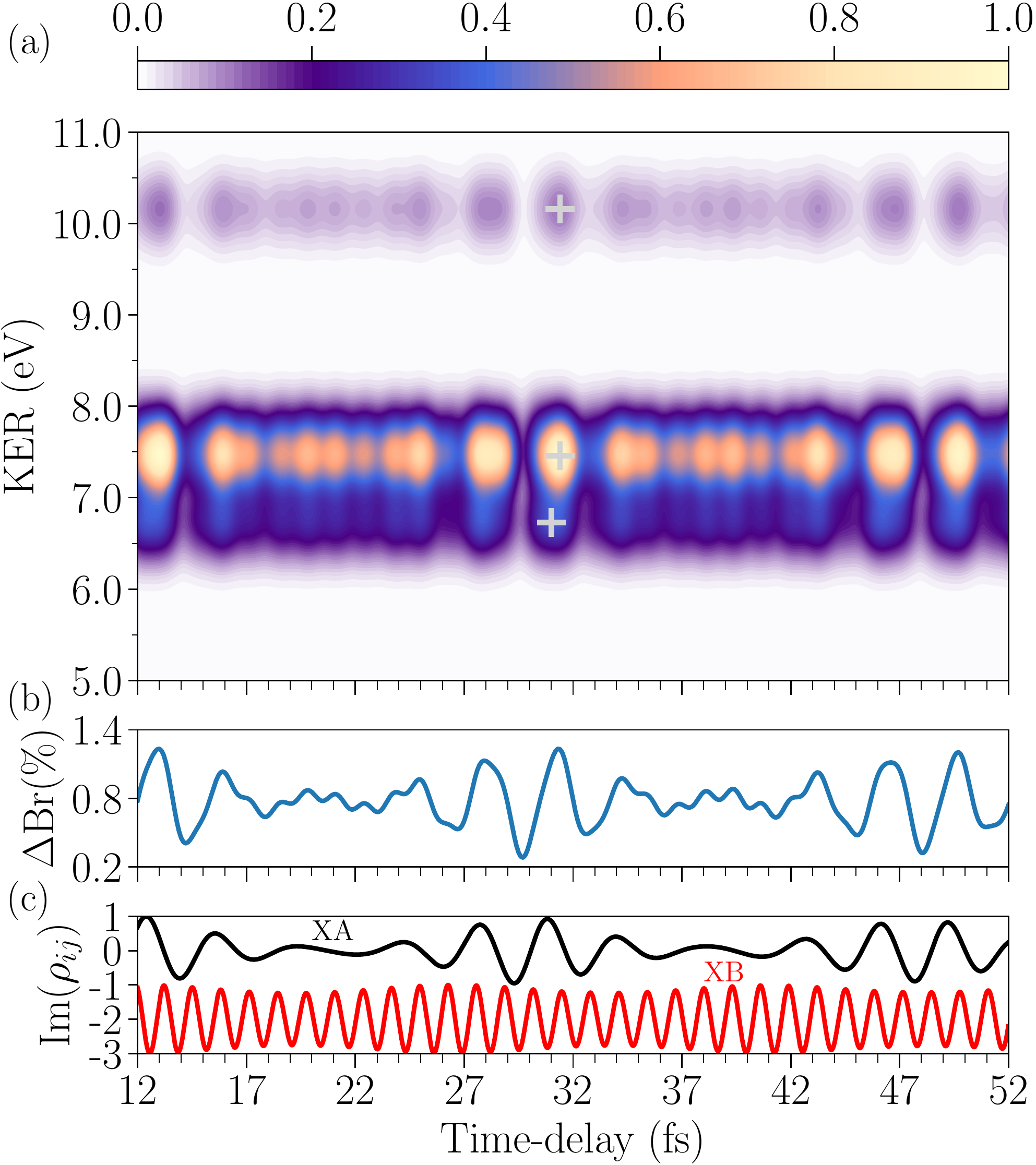} & \includegraphics[scale=0.435]{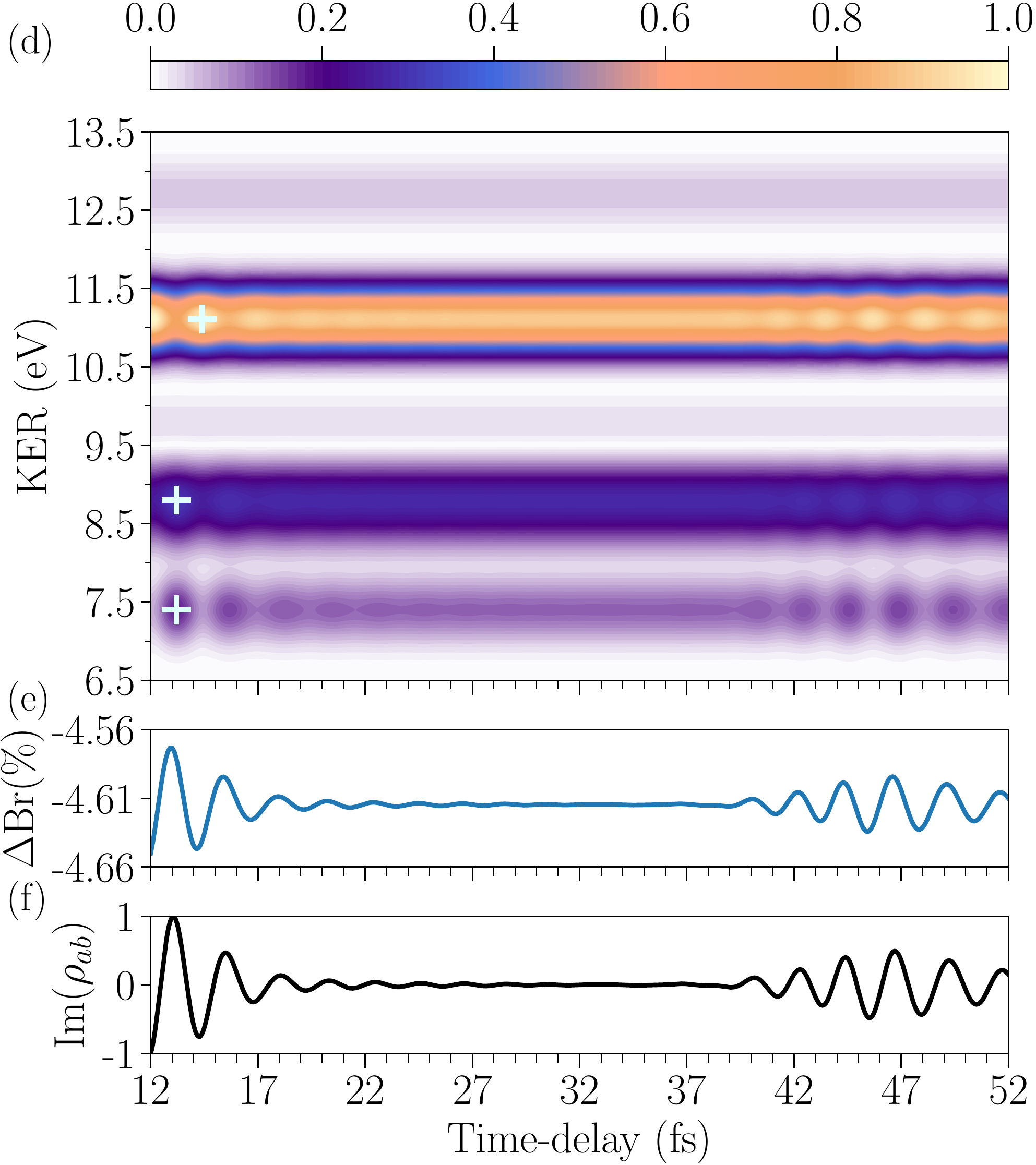}
\end{tabular}
\caption{Demonstration of the vibronic coherence dependence in the kinetic energy release (KER) spectra versus pump-probe delays. 
(a) and (d): Simulated KER spectra for $\rm N^+ + N^+$ (a) and $\rm O^+ + O^+$ (d) as a function of pump-probe delays,  subtracted by the respective probe-only signal. 
The markers show the phase differences of the beating between different KER peaks.
Note that in (d) the spectra are more negative with higher values.
Both the pump and probe pulse are linearly polarized,  with 800 nm wavelength and 6 fs pulse duration.  
The peak intensity of the pump and the probe laser are $3 \times 10^{14}$ and $1.2 \times 10^{15}$ W/cm$^2$,  respectively.  
The time-delay is defined as the time difference between the peak of the two laser pulses,  with the pump laser arriving first.  
The pump laser initiates the dynamics by simultaneously ionizing the HOMO,  HOMO-1,  and HOMO-2 of N$_2$ or O$_2$.  After some time-delays,  the vibrational motion of the ion causes changes in coherence between different electronic states.  
The probe laser then initiates subsequent tunneling ionization of the ion as well as sequential double ionization of the remaining neutral molecule,  yielding different dication states. 
(b) and (e): Branching ratio of $\rm N^+ + N^+$ and $\rm O^+ + O^+$ over their respective total dication yield as a function of pump-probe delay,  subtracted by the probe-only ratio.
(c): Imaginary part of the angular-averaged off-diagonal density matrix elements between the $X^2\Sigma_g^+$ and $A^2\Pi_u$ states (black) and the $X^2\Sigma_g^+$ and $B^2\Sigma_u^+$ states (red) of N$_2^+$.  The red curve is shifted down for better visualization.
(f): Same as (c),  but for the $a^4\Pi_u$ and $b^4\Sigma_g^-$ states of O$_2^+$. 
Due to the interference of formation pathways of the dications,  the time dependence of the density matrix elements in (c) and (f),  which represent the vibronic coherence between the ionic states,  are reflected in (a--b) and (d--e),  respectively.}
\label{fig:main}
\end{figure*}

%[Theoretical approach.  ]
%[Briefly promote the DM-SDI model.]
The DM-SDI model is based on a density matrix approach and can describe the evolution of population and coherence of different charge states due to laser couplings and tunneling ionization simultaneously~\cite{yuen2022, yuen2023}.  
To make the model simple,  we assume the nuclei of the molecule are frozen in the presence of a few-cycle IR pulse and neglects the ionized electrons such that different charge states are incoherent.
Since many-body electronic wave functions are not involved explicitly and the nuclei are frozen,  the computational cost of the model is low,  
and one can compare the calculated results under experimental conditions by averaging over the molecular orientations and the focal volume.
The predictive power of the model has been well demonstrated recently in Refs.~\cite{yuen2022,yuen2023} by reproducing main features of the KER spectra in SDI experiments of N$_2$ and O$_2$~\cite{voss2004,wu2010b}.
We refer the details of the model to our previous articles~\cite{yuen2022,yuen2023}.
Briefly,  the equations of motion for the density matrices $\rho^{(q)}$ are
\begin{align}
 \frac{d \rho^{(q)}}{d t} = -\frac{i}{\hbar}[H^{(q)},\rho^{(q)}] + \Gamma^{(q)}(t),
 \label{eq:EOM}
\end{align}
where $q=0, 1, 2$ is the charge of the molecule and $H^{(q)} = H_0^{(q)} + \vec{d}\cdot\vec{E}$,  with $H_0^{(q)}$ being the field-free Hamiltonian,  $\vec{d}$ being the dipole moment,  and $\vec{E}$ being the electric field.
The ionization rate matrices $\Gamma^{(0)}(t) = -\sum_i \rho^{(0)}(t) W^{(0)}_{i}(t)$ and $\Gamma^{(2)}_{mn}(t) = \delta_{mn} \sum_i \rho^{(1)}_{ii}(t) W^{(1)}_{n \leftarrow i}(t)$ describes the depopulation of the only neutral state and the population of the $n$th state of the dication,  where $W^{(0)}_i$ and $W^{(1)}_{n \leftarrow i}$ are the molecular orbital Ammosov-Delone-Krainov (MO-ADK) ionization rates~\cite{tong2002} from the neutral to the $i$th ionic state and from the $i$th ionic state to the $n$th state of the dication.
To properly account for the coherence between the ionic states,  we extend our previous model~\cite{yuen2022,yuen2023} by including the strong field ionization phases~\cite{pabst2016,  xue2021} and the dephasing from the depopulation of the ion as
\begin{align}
\Gamma^{(1)}_{ij}(t) &= \rho^{(0)}\sqrt{W^{(0)}_i W^{(0)}_j}\mathrm{sgn}(E)^{(2-P_i-P_j)/2}  \nonumber \\ 
&- \rho^{(1)}_{ij} \sqrt{\sum_n W^{(1)}_{n \leftarrow i}} \sqrt{\sum_n W^{(1)}_{n \leftarrow j}},
\end{align}
where $P_i$ is the parity of the ionized orbital to reach the $i$th ionic state. 
The modeling of the dephasing term is because 
$|\rho^{(1)}_{ij}| \propto \sqrt{\rho^{(1)}_{ii}\rho^{(1)}_{jj}}$ and $d\rho^{(1)}_{ii}/dt \sim -\rho^{(1)}_{ii} \sum_n W^{(1)}_{n \leftarrow i}$,  such that $\rho^{(1)}_{ij}$ should decay with the population.
%Note that the ionization phases for the dication are neglected,  since the coherence between dication states are driven by the laser.
%We note that the inclusion of ionization phase only slightly change the final results,  since the orientation averaging washes out the fine details.
%Therefore,  for simplicity,  the ionization phases for the dication are neglected.
%Comparisons between results with or without the consideration of ionization phases will be published elsewhere.
%Since the ionization phases for targets with lower symmetry are complicated in general,  it suggests one can neglect such phases,  so that our model can indeed extend to complex targets.

The simulation for the pump-probe scheme consists of these three steps: The pump,  the free propagation,  and the probe.
For simplicity in both theory and experiment,  we consider the pump and probe pulse to be a few-cycle IR pulse.
In principle,  for a better temporal resolution,  an attosecond XUV pump pulse could be used instead.

(i) For the pumping process,  we solved Eq.~\eqref{eq:EOM} at the equilibrium geometry of the neutral molecule for a 6 fs,  800 nm,  linearly polarized Gaussian pulse with a peak intensity of $3 \times 10^{14}$ W/cm$^2$ for each angle $\theta$ between the laser polarization and the molecular axis.
The initial conditions are set as $\rho^{(0)}(\theta, t_0)=1$ and $\rho^{(1)}(\theta, t_0)=\rho^{(2)}(\theta, t_0)=0$,  such that there is only neutral population before the pump pulse.
Then,  the highest occupied molecular orbital (HOMO),  HOMO-1,  and HOMO-2 of the molecule are ionized to form a superposition of ionic states ($X^2\Sigma_g^+$,  $A^2\Pi_u$,  and $B^2\Sigma_u^+$ states of N$_2^+$; $X^2\Pi_g$,  $a^4\Pi_u$,  and $b^4\Sigma_g^-$ states of O$_2^+$),   
and at the end of the pump laser ($t=t_1$),  density matrices $\rho^{(0)}(\theta, t_1)$ and $\rho^{(1)}(\theta, t_1)$ are obtained.
Since the peak intensity is rather weak,  the dication yield from the pumping process is negligible such that $\rho^{(2)}(\theta, t_1)=0$.
We set $t=0$ at the peak of the pump pulse and $t_1 = 6\,\mathrm{fs}$.

(ii) For the free propagation,  we assume that after the pump pulse,  the vibrational states of the ion are populated according to their Franck-Condon (FC) factors.
The nuclear wave function of the $i$th ionic state $|\chi_i(t)\rangle$ then evolves as $| \chi_i(t) \rangle = \sum_{v} |c_{iv}|^2 |\phi_{iv} \rangle e^{-i E_{iv} t}$,  where $|c_{iv}|^2$ is the FC factor from the neutral vibronic ground state to the $v$-level of the $i$th ionic state,  $|\phi_{iv} \rangle$ is the vibrational wave function,  and $E_{iv}$ is the vibronic energy.
To simulate the change in coherence due to the nuclear motion of the ion during the pump-probe delay,  within the FC approximation,  we model the off-diagonal elements $\rho^{(1)}_{ij}(\theta,t)$ for $t>t_1$ as~\cite{arnold2017}
\begin{align}
\rho^{(1)}_{ij}(\theta,t) = C_{ij}(\theta) \langle \chi_j(t-t_1)| \chi_i(t-t_1) \rangle, 
\label{eq:decohere}
\end{align}
where $C_{ij}(\theta)$ is a constant to match the matrix element at $t=t_1$.
%$ = \rho^{(1)}_{ij}(\theta,t_1)/\langle \chi_i(0)| \chi_j(0) \rangle$ 
Note that the nuclear overlap function $ \langle \chi_j| \chi_i \rangle$ is independent of $\theta$,  since the nuclear motion occurs in the molecular frame and is irrespective of the molecular orientations.
Details about the nuclear overlap functions can be found in Sec.  S1 of the supplemental material.

(iii) For the probing process,  we consider the probe pulse to be the same as the pump but with a peak intensity of $1.2 \times 10^{15}$ W/cm$^2$.
To account for the vibronic coherence at a later time $t$,  we solved Eq.~\eqref{eq:EOM} again for each $\theta$ but with the initial conditions $\rho^{(0)}(\theta, t)=\rho^{(0)}(\theta, t_1)$,  $\rho^{(1)}(\theta,t)$ as in Eq.~\eqref{eq:decohere},  and $\rho^{(2)}(\theta, t)=0$.
The validity of using Eqs.~\eqref{eq:EOM} and \eqref{eq:decohere} for the probing process is proved by a rigorous derivation using the fixed nuclei and FC approximation (see Sec.  S2 of the supplemental material).
Finally,  after the probe pulse,  we average the yield of each dication state over $\theta$,  assign the KER peak for each state,  and convolve the peaks with experimental energy resolution to simulate the KER spectra,  as was done in Refs.~\cite{yuen2022, yuen2023}.
Note that we neglect the focal volume effect since it does not change the qualitative behavior of the KER spectra~\cite{yuen2022,yuen2023}.

The main results of this Letter are shown in Fig.~\ref{fig:main}. 
The time-delay $\tau$,  which is defined as the time difference between the peak of the two laser pulses with the pump pulse arriving first,  begins at 12 fs in order to minimize the overlap of the pulses.  
Figure~\ref{fig:main}a and d shows the simulated KER spectra for $\rm N^+ + N^+$ and $\rm O^+ + O^+$ as a function of time-delay,  subtracted by the respective probe-only signal.
Assignment of the KER peaks for different states of the dication can be found in Refs.~\cite{yuen2022,yuen2023}.
One can see clear beatings at the highest peak of the spectra for both $\rm N_2$ (7.5 eV) and $\rm O_2$ (11 eV).
The beating for $\rm N_2$ weakens starting from $\tau$ = 16 fs (corresponding to 4 fs after the pump pulse),  but revives shortly at $\tau$ = 28 fs,  and similar patterns occur from $\tau$ = 30 to 46 fs.
Meanwhile,  the beating for $\rm O_2$ dampens from $\tau$ = 16 fs and almost vanishes during $\tau$ = 19 to 39 fs. 
At $\tau >$ 40 fs,  one can see the revival of the beating.
Note that the weaker peaks for $\rm N_2$ and $\rm O_2$ beat similarly as the main peaks,  but some with phase shifts (see the markers in Fig.~\ref{fig:main}a and d).
These phase shifts are due to different formation pathways to the dication states.
A detailed discussion can be found in Sec.  S3 of the supplemental material.

The weakening and revival of the beatings in Fig.~\ref{fig:main}a and d is due to the change of vibronic coherence at different time-delays.
If we neglect the vibrational motion during the pump-probe delays,  the beatings for both N$_2$ and O$_2$ would simply repeat the pattern from $\tau$ = 12 to 14 fs due to the infinitely long-lived coherence (see Sec.  S3 of the supplemental material).

\begin{figure}[t]
\includegraphics[scale=0.65]{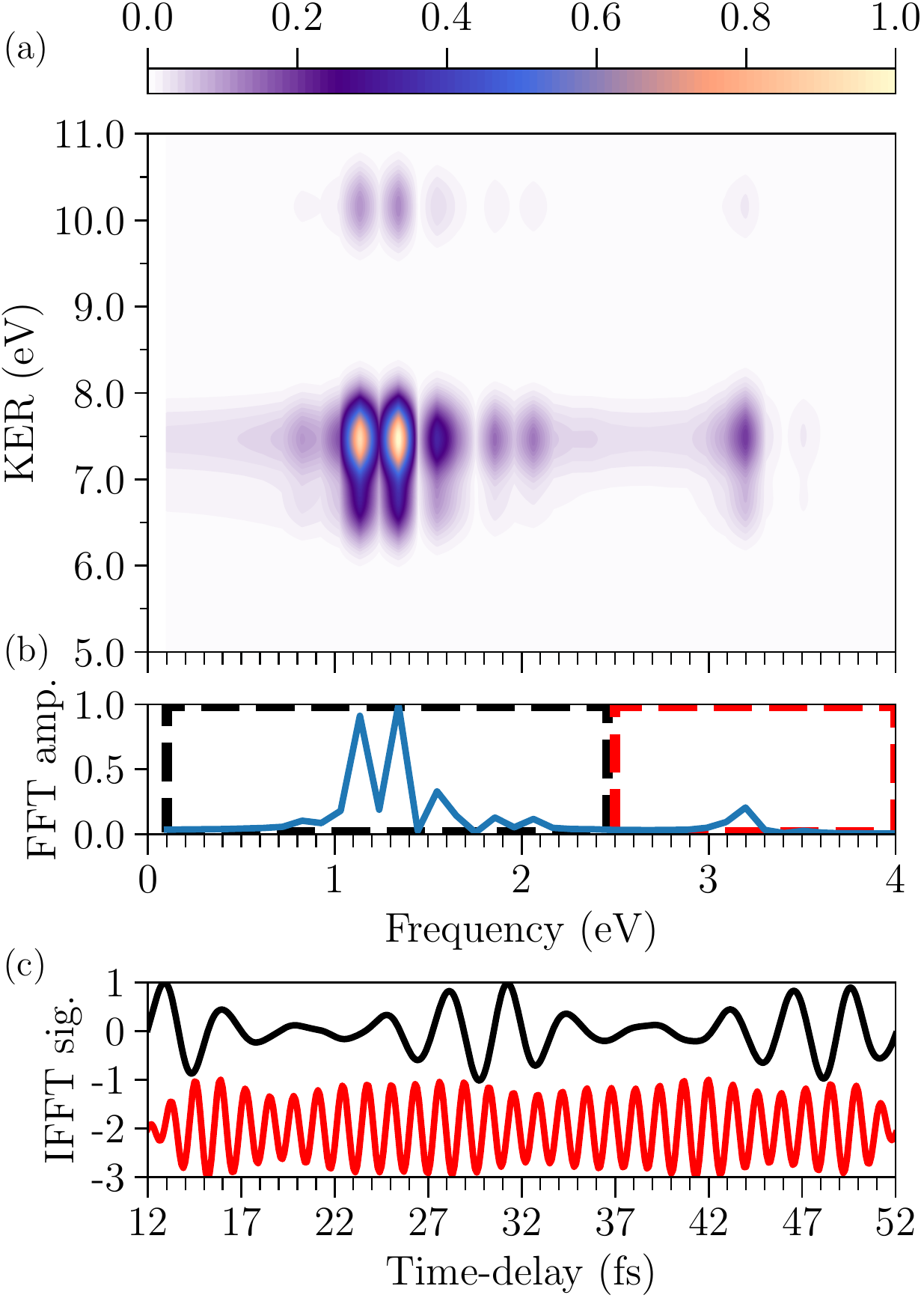}
\caption{(a--b) Fast Fourier transform (FFT) spectra of Fig.~\ref{fig:main}a and \ref{fig:main}b.  To retrieve the vibronic coherence between different pairs of states,  inverse FFT (IFFT) was performed on the signal inside the black and red boxes in (b) (and the negative counterparts),  which come from the vibronic beating of $X^2\Sigma_g^+$--$A^2\Pi_u$ and $X^2\Sigma_g^+$--$B^2\Sigma_u^+$ states of $\rm N_2^+$,  respectively.  The black and red lines in (c) are the respective IFFT signals,  with the red curve being shifted down for better visualization.
The two curves are in excellent agreement with the vibronic coherence functions in Fig.~\ref{fig:main}c.}
\label{fig:N2fft}
\end{figure}

To further understand the link between the vibronic coherence and the KER spectra,  we compare the branching ratios for dissociative dications in Fig.~\ref{fig:main}b and e with the off-diagonal density matrix elements of the ion in Fig.~\ref{fig:main}c and f at different time-delays.
The plotted branching ratios are the total yield of dissociative dications over the total yield of dications at a time-delay $\tau$,  subtracted by the probe-only ratio.
The plotted off-diagonal elements $\rho^{(1)}_{ij}(\theta,  t)$,  which represent the vibronic coherence,  are angular averaged and with $t = \tau - t_1$ (see Eq.~\eqref{eq:decohere}).
One can see that the branching ratios are closely related to the imaginary part of $\rho^{(1)}_{ij}$ for both N$_2$ and O$_2$.
Such a relationship is resulted from the equation of motion for the population of the ion (derived from Eq.~\eqref{eq:EOM}), 
\begin{align}
\frac{d\rho^{(1)}_{ii}}{dt} = 2 \sum_{l} \vec{d}_{il}\cdot\vec{E}(t)\, \mathrm{Im}\left(\rho^{(1)}_{li}(t)\right) + \Gamma^{(1)}_{ii}(t).
\end{align}
From the first term on the right,  one can see that the vibronic coherence regulates the population transfer between ionic states through the laser couplings.
Since the dication yields are proportional to the population of the intermediate ionic states,  the dication yields also depend on the vibronic coherence.

The relationship between the branching ratio and the vibronic coherence is quite different among $\rm N_2$ and $\rm O_2$.
In the case of $\rm N_2^+$,  as there are two pairs of laser coupled states,  its branching ratio is beating at two different sets of vibronic frequencies (see Fig.~\ref{fig:main}b).
Since the $X^2\Sigma_g^+$--$A^2\Pi_u$ beating ($\sim$ 1.35 eV) is in near-resonant with the 800 nm laser (1.55 eV) while the $X^2\Sigma_g^+$--$B^2\Sigma_u^+$ beating ($\sim$ 3.17 eV) is off-resonant, 
the branching ratio mostly depends on the $X^2\Sigma_g^+$--$A^2\Pi_u$ coherence.
For $\rm O_2^+$,  since only the $a^4\Pi_u$ and $b^4\Sigma_g^-$ states are coupled by the laser,  their vibronic coherence is directly imprinted to the branching ratio (see Fig.~\ref{fig:main}e).

Another interesting feature is how the coherence behaves differently for the two pairs of $\rm N_2^+$ states and for the pair of $\rm O_2^+$ states.
While the $X^2\Sigma_g^+$--$B^2\Sigma_u^+$ coherence of $\rm N_2^+$ is long-lived,  the coherence between other pairs of states of $\rm N_2^+$ and $\rm O_2^+$ dampen rapidly and revive at a later time.
From the results of $\rm N_2^+$ and $\rm O_2^+$,  one can see that the decoherence effect is highly state-dependent and system-dependent,  and the vibronic decoherence and revival~\cite{matselyukh2022} can occur even in such simple molecules.
Since the mechanism of the SDI probe should be general,  we expect similar relationships to exist for more complex molecules,  as long as the pumped states do not dissociate during pump-probe delays.

\begin{figure}[t]
\includegraphics[scale=0.65]{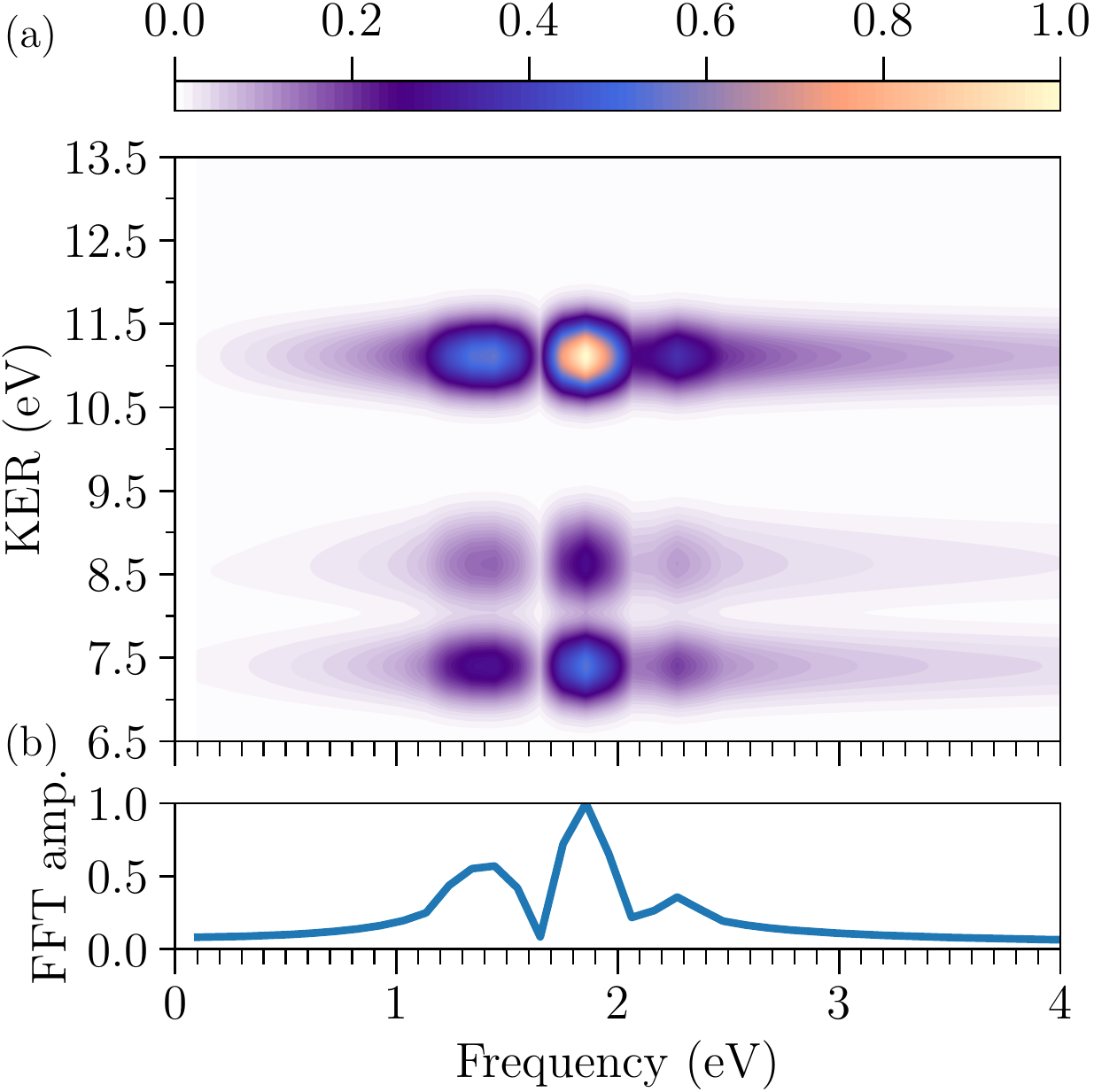}
\caption{(a--b) FFT spectra of Fig.~\ref{fig:main}d and \ref{fig:main}e.  While the $X^2\Pi_g$ state of $\rm O_2^+$ is still partially coherent to the $a^4\Pi_u$ and $b^4\Sigma_g^-$ states,  
only beating between the $a^4\Pi_u$ and $b^4\Sigma_g^-$ states are presented here,  as only this pair of states is coupled by the laser.}
\label{fig:O2fft}
\end{figure}

An important consequence of this work is that one can retrieve the vibronic coherence directly from experiments.
Since the nuclear overlap functions are independent of molecular orientation,  the retrieval makes sense even if the molecules are not aligned.
Here we show that it is possible to do the retrieval by performing a fast Fourier transform (FFT) on the spectra in Fig.~\ref{fig:main}.
Figure~\ref{fig:N2fft}a and b show the normalized amplitude of FFT spectra for Fig.~\ref{fig:main}a and b for N$_2$.
In Fig.~\ref{fig:N2fft}b,  one can see that there are two sets of beatings: One for $X^2\Sigma_g^+$--$A^2\Pi_u$ (black box) and one for $X^2\Sigma_g^+$--$B^2\Sigma_u^+$ (red box).
The $X^2\Sigma_g^+$--$A^2\Pi_u$ vibronic beatings could be four times stronger than the $X^2\Sigma_g^+$--$B^2\Sigma_u^+$ beating,  as expected from the near-resonant condition.
Since the two sets of beatings are well separated in frequency,  one can perform an inverse FFT (IFFT) for the spectra in the black and red boxes (and the same range in negative frequencies) to retrieve the vibronic coherence for the two pairs of states.
Figure~\ref{fig:N2fft}c shows the retrieved IFFT signals for N$_2$,  and one can see that both signals are in excellent agreement with Fig.~\ref{fig:main}c.
%  except near $\tau=12$ and 52 fs,  the retrieved signals for the $X^2\Sigma_g^+$--$B^2\Sigma_u^+$ beating are weaker.
%This is because the branching ratio is nearly at the zero frequency offset ($\sim 0.8 \%$) around $\tau=12$ and 52 fs,  such that one cannot separate out their contribution from the mixed signal.

Similarly,  Fig.~\ref{fig:O2fft}a and b show the vibronic beatings between the $a^4\Pi_u$ and $b^4\Sigma_g^-$ states for O$_2^+$.
But because only these two states are coupled by the laser,  an IFFT of the spectra in Fig.~\ref{fig:O2fft}b will simply reproduce Fig.~\ref{fig:main}e.
We note that similar vibronic beatings were reported in a theoretical work by Xue \textit{et al. }~\cite{xue2021} for strong field dissociation of O$_2^+$,  where the intensity of the IR pulse used is three orders of magnitude weaker than our case.

%Another use of the KER-time-delay spectra is to obtain information about the electronic configuration of the dication states for each KER peak.
%From Fig. ~\ref{fig:main}a for N$_2$,  we see that the KER peaks at 7.5 and 10 eV are beating in phase while the 6.5 eV peak has a small phase shift relative to the other two peaks.
%Meanwhile,  in Fig.~\ref{fig:main}d for O$_2$,  the beating around 7.4 and 8.8 eV are in phase with each other,  but are out of phase with the beating at 11 eV.
%Such phase shifts are due to different formation pathways to the dication states,  therefore one can deduce their electronic configuration from the phase shifts.
%Detailed discussion can be found in Sec.  S3 of the supplemental material.

In summary,  we have shown that it is feasible to use the SDI as a probe for molecular dynamics.
While the vibronic coherence in charge migration of molecules is known to be challenging to probe,
we showed that the SDI probe offers a simple and viable solution since it is driven by laser couplings between ionic states,  thereby sensitive to their coherence.
Such conclusions can be drawn from our simulations because our model takes into account of experimental conditions such as isotropic distribution of molecular orientations,  which is rarely addressed by other theoretical studies based on first-principles approaches.

The essence of our model is in the separation of the treatment of the laser-molecule interaction and the nuclear dynamics.
Because of the use of few-cycle IR pulses,  the former can be dealt with using the DM-SDI model~\cite{yuen2022, yuen2023} at fixed nuclei,  while the latter can be modeled by quantum chemistry approaches without the consideration of laser fields and molecular orientations~\cite{despre2015,golubev2020}.
Therefore,  the computational cost of the simulation is significantly reduced,  and the orientation averaging can be performed to directly compare the results with experiments. 
Consequently,  the current approach could be extended to more complex molecules.

The simplicity of the DM-SDI model also allows one to retrieve the information about the molecular dynamics from experimental observables.
In case there is no population transfer between pumped states and dissociation during pump-probe delays,  the model suggests that the vibronic coherence could be extracted directly by performing a Fourier analysis on the KER spectra or the branching ratios.
%While if there is population transfer,  other retrieval techniques could be used with the DM-SDI model to obtain the information about the evolution of population and coherence during pump-probe delays.

Finally,  we emphasize that the theory in this Letter is greatly simplified due to the use of ultrashort pulses in the pump-probe scheme.
With the recent advancement of pulse compression techniques,  it is now possible to generate few-cycle to even single-cycle IR pulses~\cite{tsai2022}.
This theoretical and experimental progress will open up numerous research opportunities for probing molecular dynamics utilizing the SDI process and the ultrashort pulses.
%Our results strongly encourage future charge migration experiments to be performed using the SDI probe.
%Once the foundation of this probing scheme is established,  it will be highly desirable to have systematic investigations into the charge migration of different types of molecules and their applications.

\begin{acknowledgments}
This work was supported by Chemical Sciences, Geosciences and Biosciences Division, Office of Basic Energy Sciences, Office of Science, U.S. Department of Energy under Grant No. DE-FG02-86ER13491.
\end{acknowledgments}


\end{document}